\documentclass[Afour,sagev,times]{sagej}

\usepackage{moreverb,url}
\usepackage{makecell}
\usepackage{algorithmic}
\usepackage{algorithm}
\usepackage{natbib}
\setcitestyle{numbers,square}

\usepackage[colorlinks,bookmarksopen,bookmarksnumbered,citecolor=red,urlcolor=red]{hyperref}

\newcommand\BibTeX{{\rmfamily B\kern-.05em \textsc{i\kern-.025em b}\kern-.08em
T\kern-.1667em\lower.7ex\hbox{E}\kern-.125emX}}

\def\volumeyear{2016}

\begin{document}

\runninghead{Smith and Wittkopf}

\title{When Mining Electric Locomotives Meet Reinforcement Learning}

\author{Ying Li\affilnum{1}, Zhencai Zhu\affilnum{1}, Xiaoqiang Li\affilnum{2}, Chunyu Yang\affilnum{3} and Hao Lu\affilnum{1}}

\affiliation{\affilnum{1}School of Mechanical and Electrical Engineering, China University of  Mining and Technology\\
\affilnum{2}School of Electrical Engineering, China University of  Mining and Technology\\
\affilnum{3}School of Information and Control Engineering, China University of Mining and Technology}

\corrauth{Zhencai Zhu, School of Mechanical and Electrical Engineering, China University of  Mining and Technology, Xuzhou 221116, China.}

\email{zhuzhencaijs@163.com}

\begin{abstract}
As the most important auxiliary transportation equipment in coal mines, mining electric locomotives are mostly operated manually at present. However, due to the complex and ever-changing coal mine environment, electric locomotive safety accidents occur frequently these years. A mining electric locomotive control method that can adapt to different complex mining environments is needed. Reinforcement Learning (RL) is concerned with how artificial agents ought to take actions in an environment so as to maximize reward, which can help achieve automatic control of mining electric locomotive. In this paper, we present how to apply RL to the autonomous control of mining electric locomotives. To achieve more precise control, we further propose an improved $\varepsilon$-greedy (IEG) algorithm which can better balance the exploration and exploitation. To verify the effectiveness of this method, a co-simulation platform for autonomous control of mining electric locomotives is built which can complete closed-loop simulation of the vehicles. The simulation results show that this method ensures the locomotives following the front vehicle safely and responding promptly in the event of sudden obstacles on the road when the vehicle in complex and uncertain coal mine environments. 
\end{abstract}


\maketitle

\section{Introduction}
In recent years, with the increasing application of auxiliary transportation equipment in coal mines, more accidents have also occurred. According to data reported on the website of the National Mine Safety Administration \cite{ref1}, there were 168 coal mine accidents nationwide in 2022, resulting in 245 deaths. Mining electric locomotive, as the main means of transportation in coal mines, undertakes the transportation of coal, gangue, equipment, materials, workers, et al. But the traditional transportation system of coal mine has the characteristics of cumbersome transfer links, large demand for personnel, complex human behavior, and backward technical equipment. There are still some dangerous factors caused by improper driving practices or by the driving environment in coal mines when mining electric locomotives are running.

In China, underground mining is commonly used as a coal mining method \cite{ref2}. The environment of coal mine roadway underground is complex, including wet and slippery ground, narrow and closed space, many road corners and insufficient light. These hostile environments would result in many hidden dangers, such as locomotive rear-end collision and crushing of pedestrians. Moreover, almost all mining electric locomotives are driven manually. Coal miners often work in a rotation of three shifts. But it is difficult to ensure that workers are always focused during the 8 hours of work. Developing the driving method of mining electric locomotives into autonomous driving can reduce the participation of workers in hazardous working conditions. In the process of designing the control algorithm for autonomous mining electric locomotives, unsafe working conditions should be fully considered as constraints for the method, which can avoid the driver's subjective thinking process being incomplete in emergency situations. 

Autonomous control of the mining electric locomotives offers a promising solution to the problems of manual control in the hazardous mining environment. The key technology of autonomous control of mining electric locomotives is to adjust the locomotives' speed dynamically under different working conditions. Many scholars have conducted research on the velocity planning of train. Yongduan Song et al. \cite{ref3} investigated the optimal operation of high speed train and established a new energy-saving strategy through dual optimization speed curve method on the basis of traction, where the high-speed train characteristics and the railway conditions are full considered. Jianwei Qu et al. \cite{ref4} proposed an energy efficient driving strategy considering regenerative braking energy, arbitrary speed limits, variable grade profiles and curve radius. The traditional speed curve optimization algorithm has the disadvantages of poor robustness and easy to fall into local optimization under extreme road conditions. Reference \cite{ref5,ref6} proposed a speed curve optimization algorithm based on improved genetic algorithm with fast convergence speed, high optimization accuracy, and good robustness. Moreover, the dynamic programming method can use practical knowledge and experience to improve the efficiency of solution. This method reflects the connections and characteristics of dynamic process evolution, so it is often applied in the design of train speed curves \cite{ref7, ref8}. Reference \cite{ref9} formulated a model  and designed an algorithm for maximizing the utilization of regenerative energy (MURE) by using the proposed approximate dynamic programming (ADP) approach to adjust the speed curve of the accelerating train. To reduce the delay rate of trains, Multi-scale Dynamic Programming was proposed to optimize the train’s regulation \cite{ref10}. 

At present, the optimization design technology for train speed has become increasingly mature. In the process of velocity planning, train operating conditions, time saving and energy-saving performance are often considered. However, the construction of railway trains is different from that of mining electric locomotives. The driving vision of railway vehicles is relatively broad, and the driving section is closed without pedestrians, so there are fewer unexpected situations compared to underground coal mines. Due to the complexity of working conditions, the autonomous control of mining electric locomotives in coal mine roadway is rarely studied. 

Reinforcement learning (RL) is a learning method that an agent learns from the interactions with the environment\cite{ref11, ref12, ref13}. Markov Decision Process (MDP) is usually used as the mathematical formalization for RL. The goal of RL is to learn an optimal policy that maps from states to actions and maximizes the discounted accumulated rewards \cite{ref14, ref15, ref16}. In the past two decades, RL has been applied in many industrial fields \cite{ref17, ref18, ref19, ref20}. Recently, RL has been applied in train operation control \cite{ref21, ref22, ref23}. We also found that adjusting the speed of mining electric locomotives dynamically can be formulated as MDP. Specifically, we can divide time into consecutive monitoring interval. At a time slot, the agent records the state (mining working conditions), takes an action (speed adjustment) and receives a reward.

Therefore, we adopt Q-learning, a basic off-policy RL algorithm that can find an optimal action selection strategy in MDP problems \cite{ref24, ref25}. Q-Learning is an algorithm of RL that solves the optimal strategy through continuous trial and error, repeated exploration, and learning without prior environmental information. This is simply an algorithm tailored to the complex and ever-changing environment of coal mines. $\varepsilon$-Greedy is the most commonly used decision-making methods for Q-learning. It has two strategies: exploration and exploitation, which can avoid itself falling into local optimum. However, the issue of balancing the relationship between exploration and exploitation urgently needs to be solved, which determines whether algorithm learning efficiency can be improved.

In order to overcome the issues of traditional algorithms applied in the different and complex environment of coal mines, we carry out research on the autonomous control of mining electric locomotives. First, we use Simpack to model the dynamics of the mining electric locomotive (Section II). Then we obtain the vehicle optimal driving/braking torque range applied through the co-simulation analysis of MATLAB/Simulink and Simpack (Section III). Furthermore, we do research on the speed planning and realize the self-learning process of the mining electric locomotive control based on RL. We propose an improved $\varepsilon$-greedy (IEG) strategy to speed up the convergence of the RL and reduce the number of iterations of RL (Section IV). To verify the effectiveness of the IEG, we build a co-simulation platform based on Simapck, MATLAB/Simulink and Python and prove that the algorithm works (Section V).

The main contributions of this paper are as follows:

(1) This paper is the first to apply Reinforcement Learning in autonomous control of mining electric locomotives . This method has changed the existing way of manually driving mining electric locomotives. We make full use of the advantage of RL to study how agents maximize the rewards in complex and uncertain environments, and realize the universality of mining electric locomotive autonomous control strategies in different coal mine roadway environments. 

(2) Compared with \cite{ref26,ref27}, this paper proposes an improved $\varepsilon$-epsilon strategy based on RL to promote the efficiency of the autonomous control algorithm. This method can enable RL to better balance the relationship between exploration and exploitation, and enhance the learning efficiency of RL.

(3) This paper builds a co-simulation platform for autonomous control and realized dynamic closed-loop simulation control of mining electric locomotives. 

\section{Dynamic Modeling of the Mining Electric Locomotive and Design of Initial Torque Range}
In this section, we build a 3D-visual dynamic model of CTY1.5/6 mining electric locomotive without considering the specific transmission details of the motor, gear and chain in the dynamic modeling process. 
First, we measure the real mining electric locomotive to obtain the characteristics and model of the vehicle body. 
Second, we create the database of wheel profile and track profile of the mining electric locomotive. The wheel-rail model is generated according to the database. Then, we establish the wheel-rail topology and the topology between the car body and the wheel set, and obtain the whole vehicle model. 
Finally, we use the complete model for simulation calculation, and we get the best range of driving/braking force, which will be used in autonomous control method of the electric locomotives.
\subsection{Modeling of the Mining Electric Locomotive Body}
Simpack is a general multibody system simulation software enabling analysts and engineers to simulate the non-linear motion of any mechanical or mechatronic system. We can use this software to obtain the dynamic characteristics of the system by building mechanical system components, hinges, constraints and other elements. We measure the CTY1.5/6 mining electric locomotive, and use Pro/E to carry out 3D modeling to obtain the rotational inertia, mass and other characteristics of vehicle components. Then we import the car body model into Simpack through STL file, and set the relevant environment and vehicle parameters. The setting parameters are shown in Table \ref{tab:table1}. 
\begin{table}[ht]
\small\sf\centering
\caption{Model Parameters of the CTY1.5/6 Mining Electric Locomotive.\label{tab:table1}}
\begin{tabular}{lll}
\toprule
Name&Unit&Value\\
\midrule
\texttt{Adhesive weight } & t & 1.5\\
\texttt{Axle load} & t & 0.625\\
\texttt{Length} & mm & 1800\\
\texttt{Width} & mm & 800\\
\texttt{Height} & mm & 1550\\
\texttt{The type of bearing} & &22311\\
\texttt{Gauge} & mm & 600\\
\texttt{Wheelbase} & mm & 600\\
\texttt{Nominal wheel radius} & mm & 193\\
\texttt{Wheel profile} &  & GB4695-84A\\
\texttt{\makecell[l]{Minimum radius of\\ curvature negotiable}} & m & 4\\
\texttt{Braking mode} &  & \makecell[c]{Electrical braking, \\ Hydraulic braking}\\
\texttt{Hourly tractive force} & kN & 3.2\\
\texttt{Maximum tractive force} & kN & 4.65\\
\texttt{Hourly speed} & km/h & 6.5\\
\texttt{Maximum speed} & km/h & 10.2\\
\bottomrule
\end{tabular}
\end{table}

\subsection{Modeling of Wheel/rail Tread of the Mining Electric Locomotive}
Chinese standard 12kg/m track (Fig. \ref{fig:1}(a)) is compatible with CTY1.5/6 electric locomotive. The parameters of track modeling are shown in the Table \ref{tab:table2}. We use MATLAB to compile the data coordinates of rail and wheel profile, and write the data into PRR and PRW files. Then we import the files into Simpack database. The corresponding model is automatically generated through Simpack.

\begin{table}[ht]
\small\sf\centering
\caption{Track Model Parameter.\label{tab:table2}}
\begin{tabular}{lll}
\toprule
Name&Unit&Value\\
\midrule
\texttt{Length of No. 1 route} & m & 50\\
\texttt{Length of No. 2 route} & m & 50\\
\texttt{Curve radius of No. 2 route} & m & 220.48\\
\texttt{Superelevation of No. 2 route} & mm & 3.8\\
\texttt{Length of No. 3 straight line} & m & 50\\
\texttt{Track quality} & kg/m & 12\\
\bottomrule
\end{tabular}
\end{table}

The schematic diagram of GB4695-84A wheel profile is shown in Fig. \ref{fig:1}(b), in which the horizontal and vertical coordinates represent the y-axis and z-axis of the mining electric locomotive body coordinates respectively. The origin of the coordinate system is the intersection point of the wheel rolling circle and the profile.

\subsection{Modeling of the Mining Electric Locomotive System}
\begin{figure*}[!t]
\centering
\includegraphics[width=\textwidth]{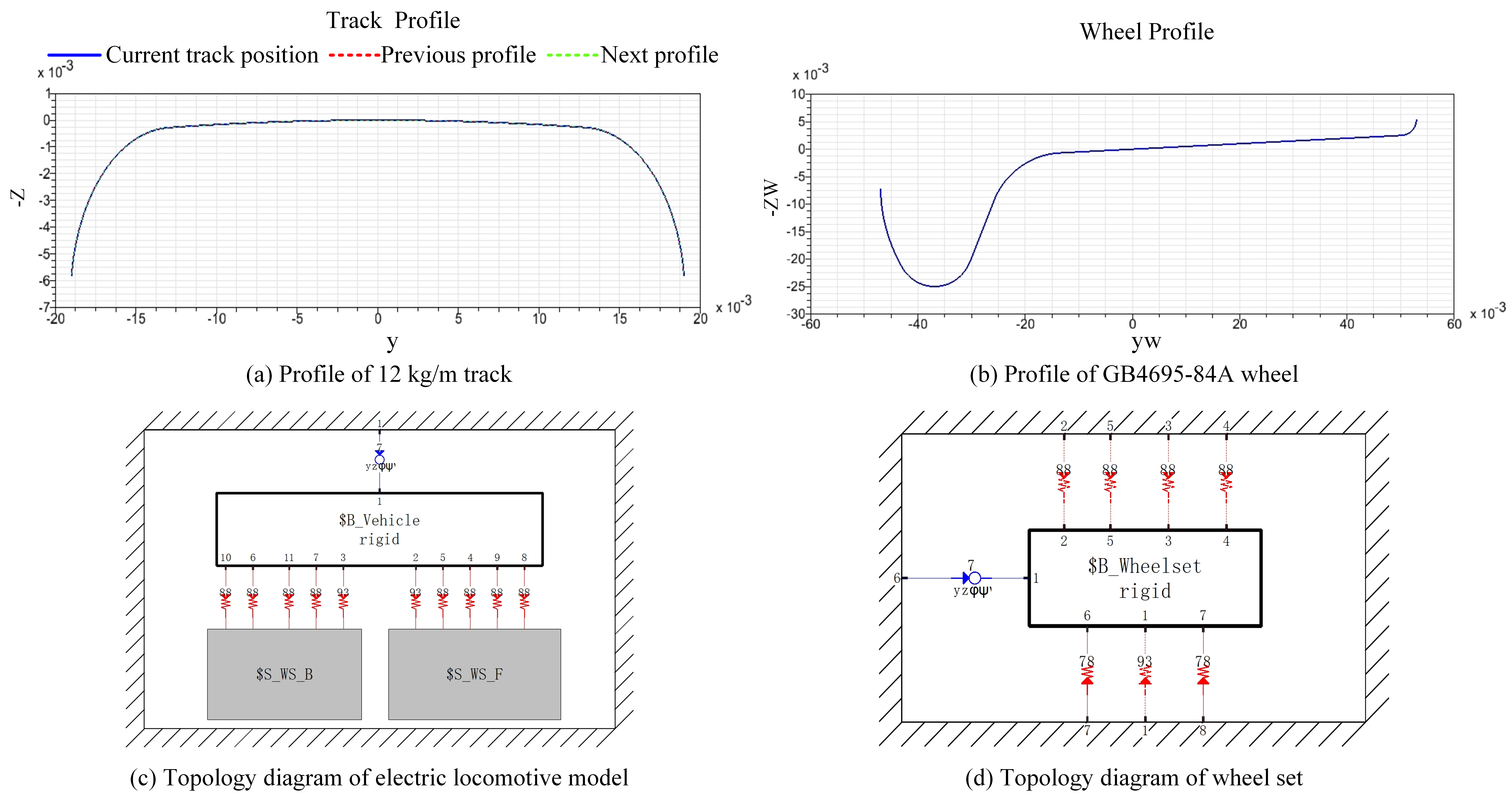}
\caption{Details of the CTY1.5/6 electric locomotive model construction.}
\label{fig:1}
\end{figure*}

\begin{figure}[!t]
\centering
\includegraphics[width=\linewidth]{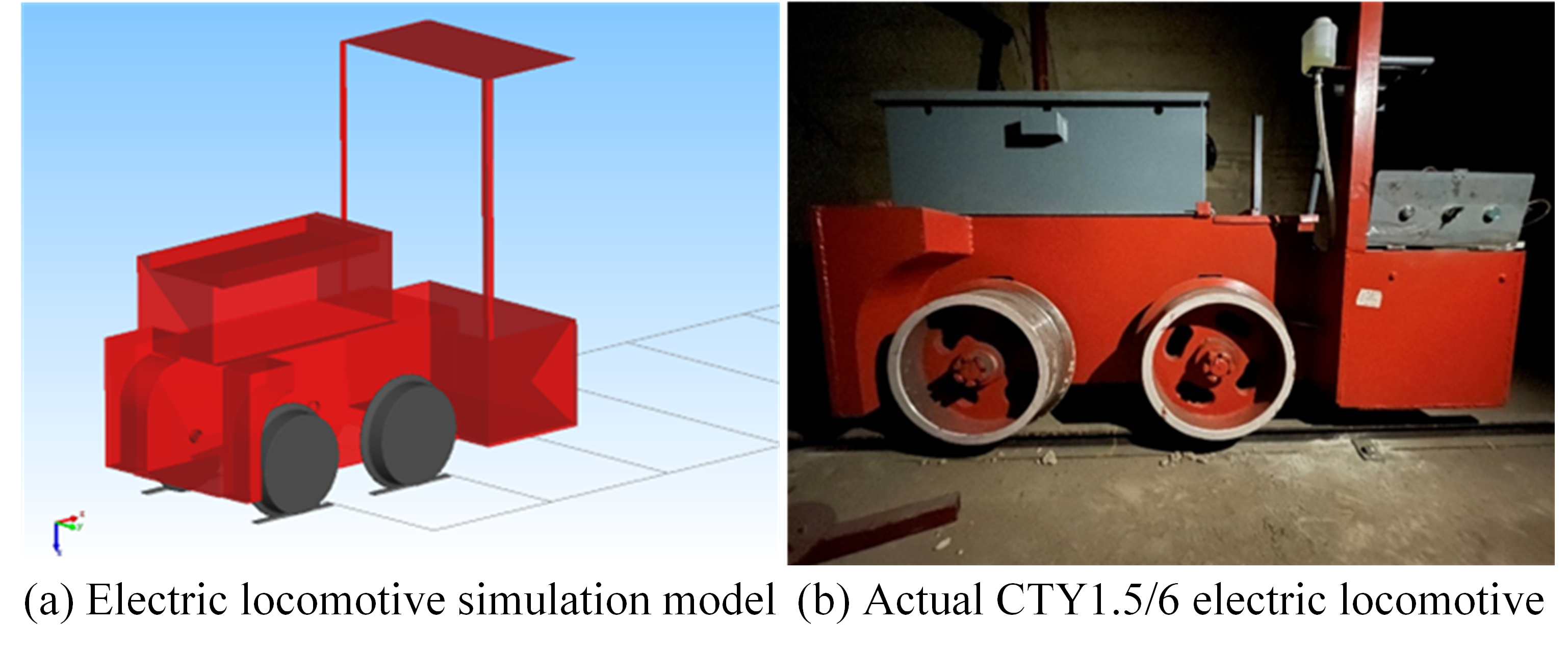}
\caption{Comparison of the mining electric locomotive model and actual vehicle.}
\label{fig:2}
\end{figure}

The influence of speed on dynamic performance of mining electric locomotives is the focus of this paper. Therefore, we focus on studying the motion characteristics of the car body and wheel pairs in modeling, without considering the power transmission between motors, gears, and chains. In Fig. \ref{fig:1}(c), we create the model composed of vehicle body and front and rear wheel sets WS\_F, WS\_B. We use No.7 hinge to represent the motion of the front and rear wheel pairs with respect to the earth. In Fig. \ref{fig:1}(d), the front and rear wheel sets are articulated with No.7 hinge. With reference to real electric locomotives, we adopt No.88 rolling bearing to connect the vehicle body and wheel sets. The comparison between the CTY1.5/6 electric locomotive model and the actual vehicle is shown in Fig. \ref{fig:2}.

\subsection{Design of Initial Torque Range Based on the Mining Electric Locomotive Model}
\begin{figure}[!t]
\centering
\includegraphics[width=\linewidth]{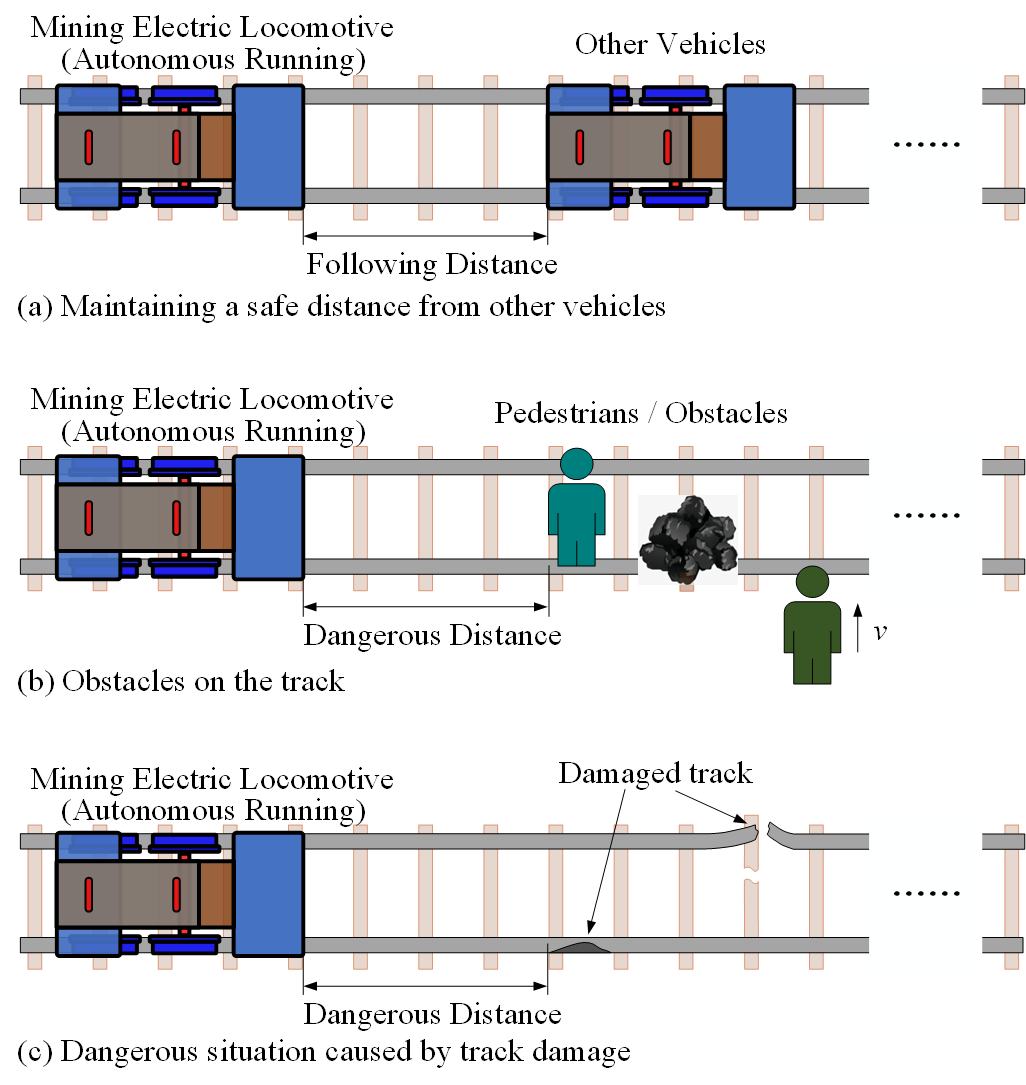}
\caption{Operating conditions of mining electric locomotives.}
\label{fig:3}
\end{figure}

As the main auxiliary transportation tool in the coal mine roadway, the mining electric locomotive is greatly affected by the humid, muddy and other harsh environment during driving. The working conditions include but are not limited to maintaining a relatively safe distance from the vehicle in front (Fig. \ref{fig:3}(a)), encountering obstacles (Fig. \ref{fig:3}(b)) or track damage (Fig. \ref{fig:3}(c)). When there are pedestrians or obstacles, or the track is damaged by heavy waste rock or other mining transport vehicles, mining electric locomotives can't bypass to avoid them limited by constraints between wheel sets and tracks, but only take emergency braking. Wet and slippery track surface may lead to insufficient wheel/rail adhesion of the mining electric locomotive. When the vehicle is in traction condition and the traction force is greater than the wheel/rail adhesion, the wheel will spin. When the vehicle is in braking condition and the braking force is greater than the adhesion, the wheels will slip. The occurrence of these situation will cause the wheel tread and rail surface to be scratched, which will seriously affect the safety and stability of the vehicle. Therefore, the most critical control factor for the safe running of the mining electric locomotive on the track is the value of the driving/braking torque applied on the axle. In this way, we can directly control the driving acceleration and speed of the vehicle by controlling the torque to avoid driving accidents such as vehicle braking failure and rollover. \\

\begin{figure}[!t]
\centering
\includegraphics[width=\linewidth]{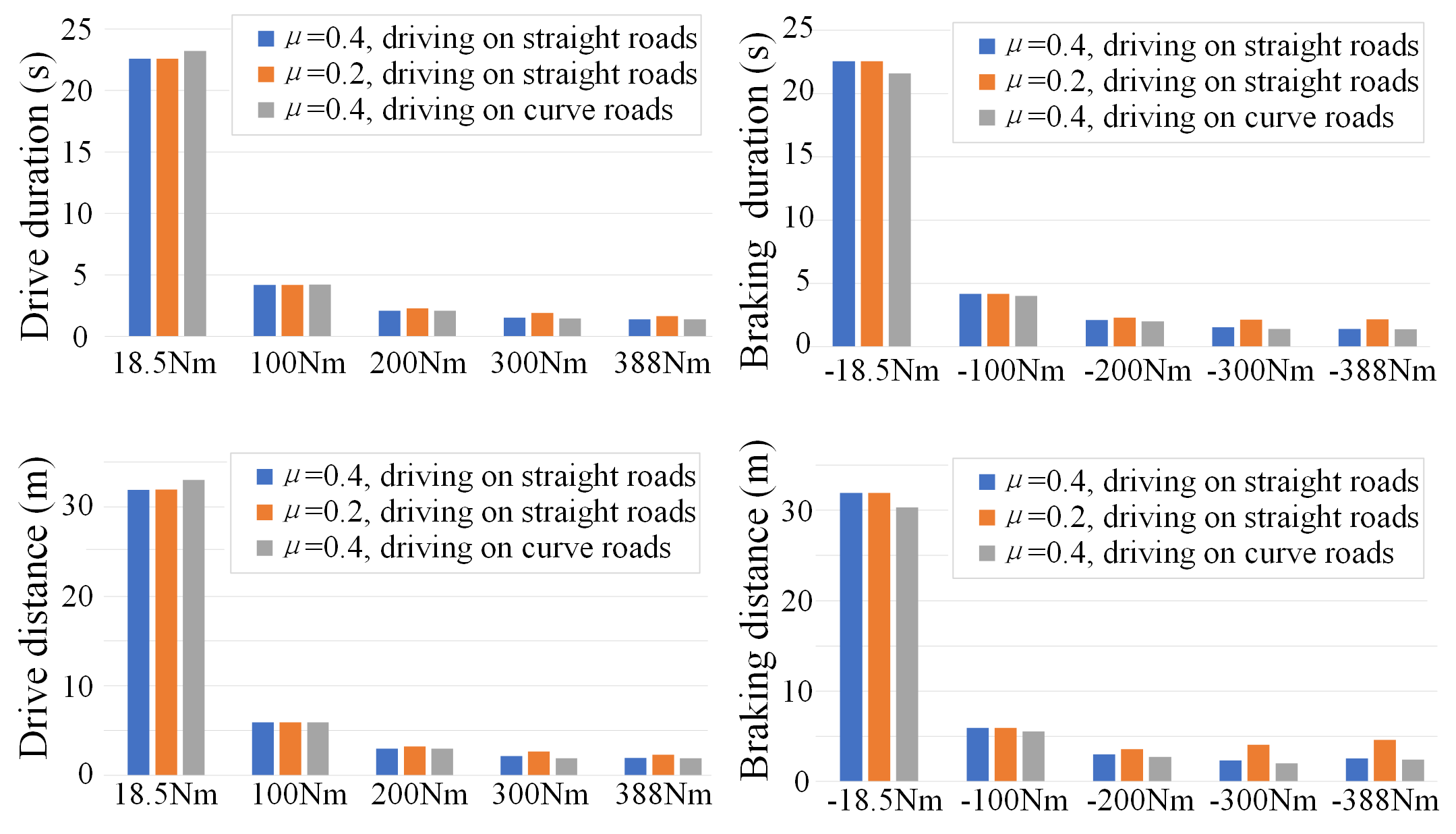}
\caption{Comparison of mining electric locomotive driving/braking results under different conditions.}
\label{fig:4}
\end{figure}

Based on the above simulation model, we obtained the driving/braking process of the vehicle under the curved/straight track and different wheel-rail friction conditions. The friction coefficient of the rail surface is influenced by different factors. Under lubrication or friction control conditions, the friction coefficient varies between 0.05 and 0.35. When the rail is dry, the friction coefficient is about 0.4. We set the wheel rail static friction coefficient in the Simpack model to 0.2 and 0.4 respectively, and conduct comparative studies on them.Fig. \ref{fig:4} shows the comparison of mining electric locomotive driving/braking results under different conditions. At larger driving/braking torque, the duration and distance of vehicle driving/braking when the static friction coefficient $\mu$ between wheel and rail is 0.2 is larger than when $\mu$ is 0.4.

\begin{figure*}[ht]
\centering
\includegraphics[width=\textwidth]{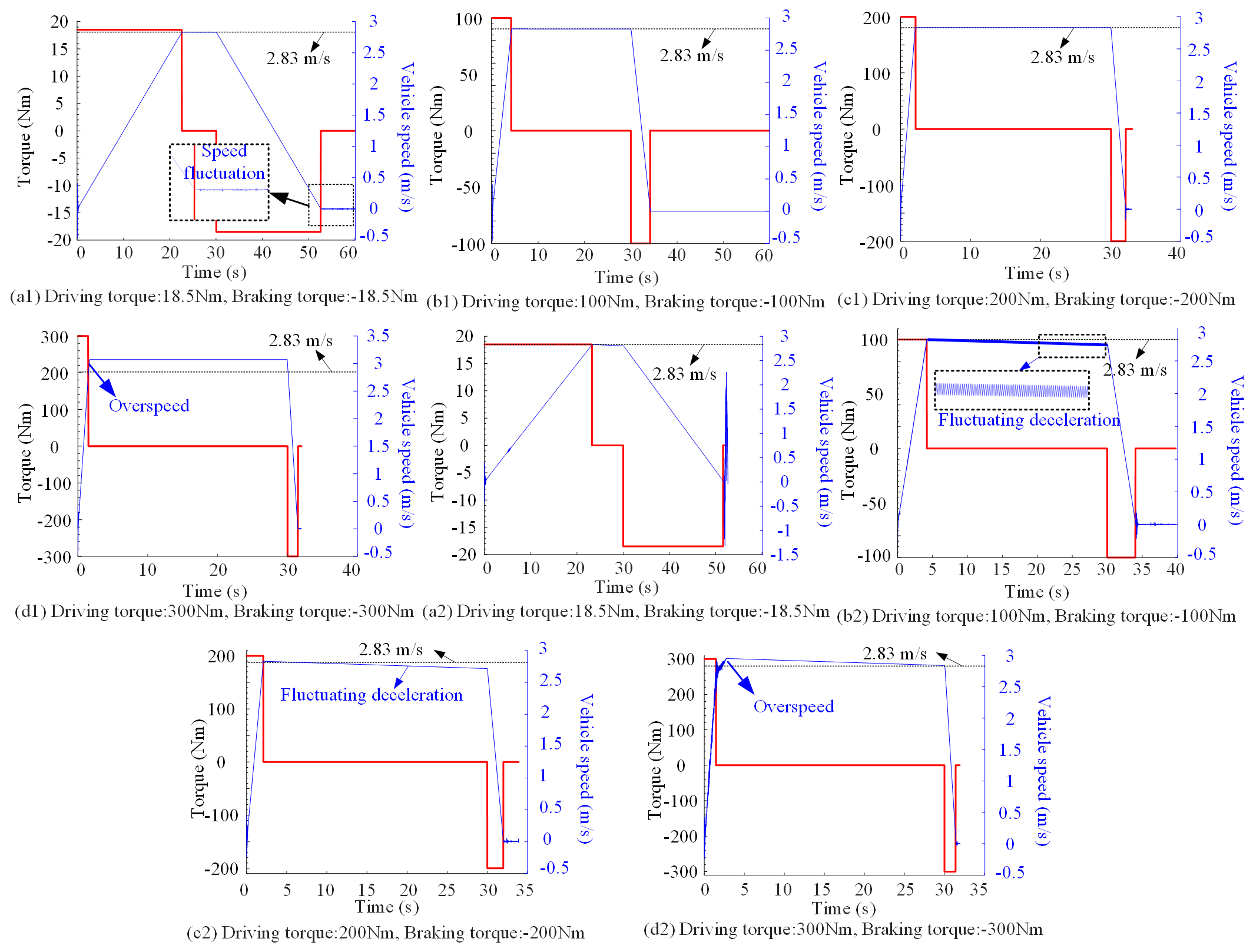}
\caption{The speed of electric locomotive under different applied force conditions.}
\label{fig:5}
\end{figure*}

When the static friction coefficient is 0.4, the running speed of the mining electric locomotive on the curve/straight track under different driving/braking torque is shown in Fig. \ref{fig:5}. In the straight track (in Fig. \ref{fig:5}(a1-d1)), when the driving torque is 18.5Nm, 100Nm, and 200Nm, the CTY1.5/6 electric locomotive can maintain the set maximum speed. When the driving torque is 300Nm and 388Nm, the vehicle still accelerates for some time without applying the driving torque due to the driving inertia. When the braking torque is -18.5Nm, -100Nm, -200Nm, -300Nm, the CTY1.5/6 electric locomotive can successfully stop near the end point. When the braking torque is -388Nm, the vehicle will be driven in reverse at a constant low speed after the vehicle slow down to 0m/s. When the vehicle is driving in the curved track(in Fig. \ref{fig:5} (a2-d2)), the driving time and distance of the vehicle to reach the maximum speed is significantly longer than the driving time in the straight track driven by the same driving torque of the axle. Moreover, when the vehicle is not driven by the driving torque, its speed fluctuates greatly, and it cannot maintain the maximum speed and uniform speed. Therefore, the initial braking speed is smaller, and the braking time and distance are significantly reduced compared with driving on the straight track.

The optimal driving/braking torque applied on the front and rear axles of CTY1.5/6 mining electric locomotive should be controlled within -300Nm - 300Nm to achieve the optimal driving/braking efficiency. 

\section{Autonomous Control Method}
We aim to address the issues of manual mining electric locomotive control with an effective automatic approach.
The proposed approach should take the right action timely and adjust the speed smoothly when running on the different mine roadway with many turns, narrow sections and emergencies.
Reinforcement Learning has great potential to dynamically adjust the speed of the mining electric locomotives.  
Because RL is suitable for making decisions in uncertain and complex driving environments. 
The autonomous control method based on RL can replace the existing manual control way of electric locomotives, and improve the safety of auxiliary transportation in coal mines. 
Thus, we propose a RL-based autonomous control method of mining electric locomotives. 
\subsection{RL and Q-Learning Algorithm}
RL can be used to imitate the learning principles of humans and animals. As shown in Fig. \ref{fig:6} (a), the environment provides the agent with the current state \textit{$S_{t}$} and the effect of the previous moment on the current moment \textit{t}, i.e., the gain \textit{$R_{t}$} (the initial moment can be assumed to be \textit{$R_{0}$} = 0). The agent picks the action \textit{$A_{t}$} based on the current state \textit{$S_{t}$} given by the environment and the gain \textit{$R_{t}$}. The action influences the state of the environment \textit{$S_{t+1}$} at the next moment \cite{ref29}.

As a trial-and-error learning method, by repeating the above interaction process between the learner and the environment, the learner eventually finds the state-to-action mapping that maximizes benefits  \textit{$R_{t}$} over the cumulative sum of time \textit{$G_{t}$}, i.e., the optimal control strategy \cite{ref28}.
\begin{equation}
\label{deqn_1}
G_{t} = \sum_{i=0}^{T-i-1}\gamma^{i} R_{t+i+1}.
\end{equation}
where \textit{t} is the current running step, \textit{T} is the total number of running steps, and $\gamma$ is discount factor.

Q-Learning is an RL strategy, which is a typical offline learning algorithm of model-independent offline strategy, and uses reasonable strategies to generate operations. It can learn to obtain another optimal Q-function according to the next state \textit{$S_{t+1}$} and the reward obtained from the interaction between the action and the environment \cite{ref30}. The Q-function \textit{$Q(S_{t},A_{t})$} is defined to estimate the action value function \textit{$q_\pi(s,a)$} for all executable actions \textit{$a$} in any state \textit{$s$} when the policy $\pi$ is taken. This algorithm updates \textit{$Q(S_{t},A_{t})$} using action that maximizes the value of \textit{$Q(S_{t+1}, a)$}, and the update formula is as follows.
\begin{equation}
\label{deqn_2}
\begin{split}
Q(S_{t}, A_{t}) & \gets Q(S_{t},A_{t})+\alpha[R_{t+1} \\ & 
+\gamma {\max}{Q(S_{t+1},a)}-Q(S_{t},A_{t})]   
\end{split}
\end{equation}
where $\alpha$ is learning rate \cite{ref31}.

The Q value is stored discretely in the Q-table, which is expressed as the probability matrix of which action to choose in the corresponding state. The Q-table can be understood as the memory of agent after learning in Q-Learning. The flow of Q-Learning is shown in Algorithm \ref{alg:alg1}.

\begin{algorithm}[ht]
\caption{Q-Learning}\label{alg:alg1}
\begin{algorithmic}
\STATE 
\STATE {\textbf{Initialize:} }$Q(S,A)=0$
\STATE {\textbf{Repeat} (for each episode):}
\STATE \hspace{0.5cm}{\textbf{Initialize:}  state }$S$
\STATE \hspace{0.5cm}{Repeat (for each step of episode):}
\STATE \hspace{1cm}{Choose action} $A_{t}$ {in state} $S_{t}$ {according to}
\STATE \hspace{1cm}{\quad Q-function (e.g $\varepsilon$-greedy)}
\STATE \hspace{1cm}{Observe the reward} $R_{t}$ {and the state} $S_{t+1}$ 
\STATE \hspace{1cm} {\quad at the next moment}
\STATE \hspace{1cm}{Update Q value and state:}
\STATE \hspace{1cm}$Q(S_{t}, A_{t}) \gets Q(S_{t},A_{t}) +\alpha[R_{t+1}$
\STATE \hspace{1cm}$\qquad \qquad  +\gamma {\max}{Q(S_{t+1},a)}-Q(S_{t},A_{t})]$
\STATE \hspace{1cm}$S_{t+1} \gets S_{t}$ {;} $t \gets t+1$
\STATE \hspace{0.5cm}{Until the state} $S$ {is terminal.}
\end{algorithmic}
\label{alg1}
\end{algorithm}

\subsection{Improved \texorpdfstring{$\varepsilon$}{}-Greedy Strategy}
$\varepsilon$-Greedy is a commonly RL method to balance exploration and exploitation by choosing between exploration and exploitation randomly \cite{ref32}. The $\varepsilon$-greedy, where $\varepsilon$ refers to the probability to explore, exploits most of the time with a small chance of exploring. $\varepsilon$-Greedy embodies the basic idea of Q-Learning algorithm to select actions. The selection of actions depends on greedy strategy, that is, to select the action that makes the current reward the largest. However, if the action is selected according to the greedy strategy at every moment, the global optimal strategy may not be obtained, because the global optimal strategy has a certain probability to contain some actions that are not the current optimal solution. Thus, the choice of action needs to balance the relationship between exploration and exploitation. The agent explores whether other solutions that are not the current optimal solution may be the component values of the optimal strategy, and selects the current optimal solution based on the greedy strategy. Therefore, $\varepsilon$-greedy algorithm is derived.
\begin{equation}
\label{deqn_3}
\begin{split}
a_{t}=
\begin{cases}
{\arg}{\max\limits_{a}} Q(S_{t+1},a), for\quad probability \quad 1- \varepsilon_{t}\\
random\quad from \quad A_{t}, for \quad probability \quad \varepsilon_{t}
\end{cases}
\end{split}
\end{equation}

\begin{figure}[!t]
\centering
\includegraphics[width=\linewidth]{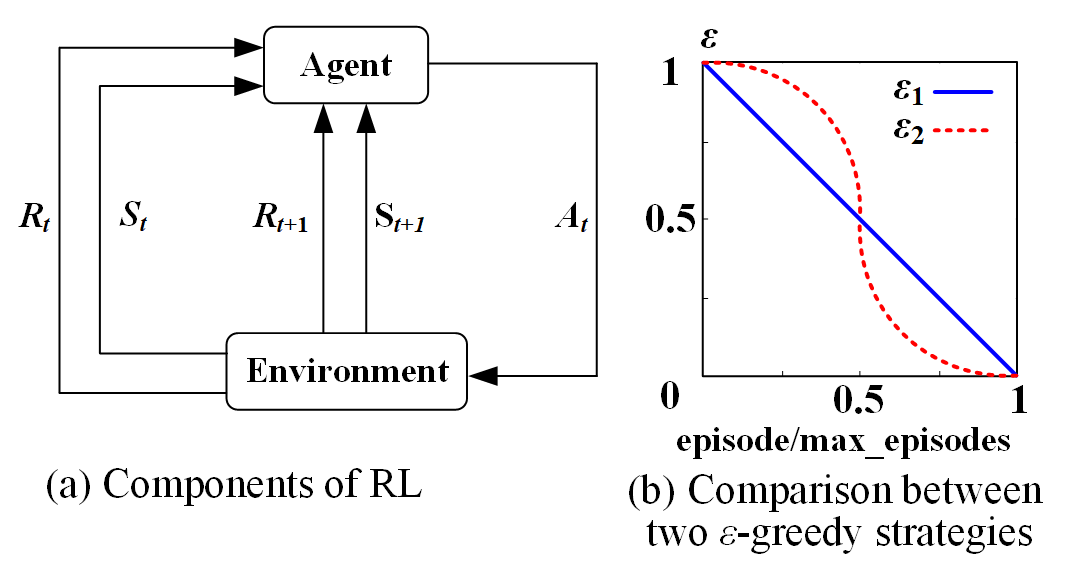}
\caption{RL and $\varepsilon$-greedy.}
\label{fig:6}
\end{figure}

For traditional strategies, too large exploration rate will lead to slow convergence or divergence of the algorithm, while too small exploration rate is not conducive to obtaining the optimal strategy. Generally, when dealing with practical problems, researchers \cite{ref26,ref27} change the original control strategy of $\varepsilon$ as a fixed value to the method in which the control parameters decrease with the number of iterations (as shown in Fig. \ref{fig:6}(b)). The purpose of this is to make the agent obtain higher exploration rate in the early stage of the RL, and rely more on experience value in the later. However, this linear decreasing method of parameters will also lead to the agent's insufficient use of empirical value in the later stage of training. Worse, the training result of the agent is not convergent, and the best possible training effect cannot be achieved under a certain number of iterations.

In order to overcome this defect and achieve better training results within a certain number of iterations, we propose an IEG strategy using the method of making parameter $\varepsilon$ nonlinear change. To explore the possibility of action comprehensively in the early stage of iteration, we design a higher exploration rate than the method of $\varepsilon_{1}$. We also design a higher utilization rate at the later stage of the iteration to ensure the reliability of the agent training. In general, to make the reward of autonomous operation control of the mining electric locomotive based on RL converge as soon as possible, we propose an improved $\varepsilon$ value taking method based on the $\varepsilon_{1}$, the formula is as follows:
\begin{equation}
\label{deqn_4}
\begin{split}
\varepsilon_{1}=(\varepsilon_{\text{initial}}-\varepsilon_{\text{final}}) \cdot (1-episode/max\_episodes)
\end{split}
\end{equation}

When (episode/max\_episodes) $\in$ [0, 0.5],
\begin{equation}
\label{deqn_5}
\begin{split}
\varepsilon_{2} = & (\varepsilon_{\text{initial}}-\varepsilon_{\text{final}}) \cdot (0.5+ \\ &
\sqrt{0.25-(episode/max\_episodes)^{2}})
\end{split}
\end{equation}

When (episode/max\_episodes) $\in$ (0.5, 1],
\begin{equation}
\label{deqn_6}
\begin{split}
\varepsilon_{2} = & (\varepsilon_{\text{initial}}-\varepsilon_{\text{final}}) \cdot (0.5- \\ &
\sqrt{0.25-(1-episode/max\_episodes)^{2}})
\end{split}
\end{equation}

\subsection{Autonomous Control Method of the Mining Electric Locomotive}
When solving the autonomous control problem of mining electric locomotives, we should first deal with how to transform the actual vehicle autonomous control problem into MDP. We take the electric locomotive as an agent. When the driving conditions of the electric locomotive are different, the intelligent agent selects reasonable actions according to the current state. Since the actual vehicle driving process is continuous, we discretize the driving process every 0.01s to control the vehicle running speed in real time, so as to ensure that the agent can quickly obtain the latest environmental information and make the best action decision.

(1) State

We propose to divide the vehicle driving state into 16 states as shown in Table \ref{tab:table3}, according to whether the vehicle runs in the curved/straight track, whether it encounters obstacles, whether it reaches the maximum driving speed and whether it reaches the destination.

\begin{table}[ht]
\small\sf\centering
\caption{State Design of the Mining Electric Locomotive Autonomous Control Based on RL\label{tab:table3}.}
\begin{tabular}{ll}
\toprule
State&Meaning\\
\midrule
\texttt{begin} & Electric locomotive start\\\
\texttt{to\_the\_end} & \makecell[l]{Electric locomotive reaches \\ the destination}\\
\texttt{obstacle\_stop} & \makecell[l]{Stop when encountering \\ obstacles}\\
\texttt{c\_within\_obstacle} & \makecell[l]{Close to the vehicle in \\ front on curves}\\
\texttt{max\_spd\_c\_no\_obstacle} & \makecell[l]{The maximum set speed is 
 \\ reached on curves}\\
\texttt{over\_spd\_c\_no\_obstacle} & Overspeed on curves\\
\texttt{below\_spd\_c\_no\_obstacle} & \makecell[l]{The maximum set speed is \\ not reached on curves}\\
\texttt{l\_within\_obstacle} & \makecell[l]{Keep close to the obstacle \\ in front when driving straight}\\
\texttt{near\_to\_the\_end\_brake} & \makecell[l]{The speed is greater than 0 \\ when approaching the \\terminal}\\
\texttt{near\_to\_the\_end\_drive} & \makecell[l]{The speed is less than 0 \\ when approaching the \\terminal}\\
\texttt{max\_spd\_l\_to\_c} & \makecell[l]{The maximum speed allowed \\in the curve is reached when \\ preparing to turn}\\
\texttt{over\_spd\_l\_to\_c} & \makecell[l]{Overspeed when preparing to \\turn}\\
\texttt{below\_spd\_l\_to\_c} & \makecell[l]{The maximum speed allowed \\in the curve is not reached \\ when the vehicle is preparing \\ to turn}\\
\texttt{max\_spd\_l\_no\_obstacle} & \makecell[l]{The maximum set speed is \\ reached on the straight track}\\
\texttt{over\_spd\_l\_no\_obstacle} & Overspeed on the straight\\
\texttt{below\_spd\_l\_no\_obstacle} & \makecell[l]{The maximum set speed is  \\ not reached on the straight \\ track}\\
\bottomrule
\end{tabular}
\end{table}

(2) Action

\begin{figure*}[ht]
\centering
\includegraphics[width=\textwidth]{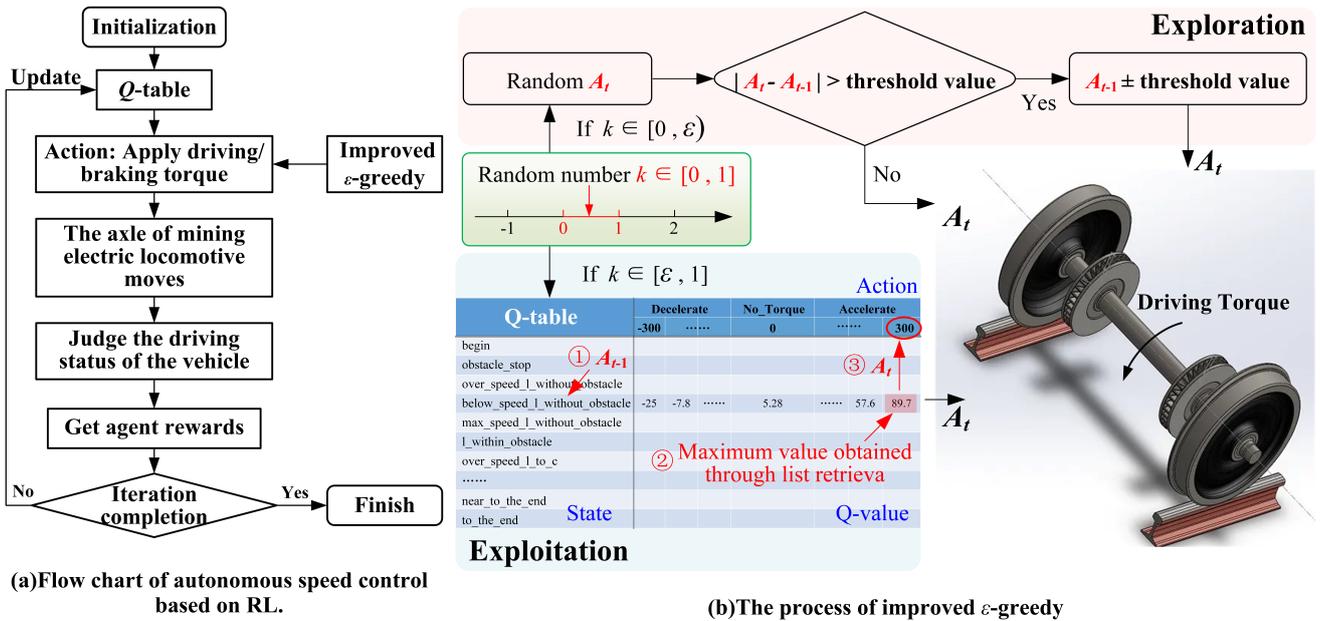}
\caption{Improved $\varepsilon$-greedy (IEG).}
\label{fig:7}
\end{figure*}

After the analysis of the drive/braking performance of the CTY1.5/6 electric locomotive, we found that the optimal drive/braking torque range of the vehicle is between -300Nm and 300Nm. Therefore, we set the agent action to an integer within [-300, 300]. In Fig. \ref{fig:7}(a), we put forward the process framework of autonomous control for the mining electric locomotive on RL, and give the concrete implementation steps of the IEG in Fig.\ref{fig:7}(b). The action decision process is as follows.

\textbf{Step 1}$\colon$ According to the IEG, the agent obtains the driving action randomly or by virtue of experience value.

\textbf{Step 2}$\colon$ In order to ensure the smoothness of the vehicle's running speed and prevent the vehicle from running fault due to sudden too small/too large speed, the agent makes a judgment on whether the difference between $A_{t}$ and $A_{t-1}$ exceeds the threshold. 

\textbf{Step 3}$\colon$ If the threshold value is exceeded, the agent will choose whether to accelerate or decelerate based on Step 1. The value of acceleration or deceleration torque is determined based on the action taken at the last moment $t$-1. If the threshold is not exceeded, the agent directly executes the action of Step 1.

(3) Reward function

We set different reward values based on the actions of the agent in different states. In the process of designing rewards, we consider the efficiency of vehicle running speed and whether the driving speed conforms to the safety regulations.
\begin{equation}
\label{deqn_7}
\begin{split}
reward=c*(v_{\text{record}}-v_{\text{current}})
\end{split}
\end{equation}
where, $v_{\text{record}}$ is the feedback value of vehicle driving speed in the last sampling interval, $v_{\text{current}}$ is the feedback value of vehicle driving speed in this sampling time, and $c$ is the reward function coefficient. The positive or negative of the formula is determined by whether the vehicle is speeding up.

\section{Simulation Verification and Evaluation}
In this section, we evaluate the feasibility and performance of IEG algorithm on our co-simulation platform. In the simulation, we simulate the traveling speed of mining electric locomotives on the straight and curved track. We also verify whether the vehicle can start and brake safely at the destination through co-simulation platform, and maintain a safe distance from dynamic obstacles. Finally, we obtain the simulation results and analyze them.
\subsection{Simulation Platform}
\begin{figure}[ht]
\centering
\includegraphics[width=\linewidth]{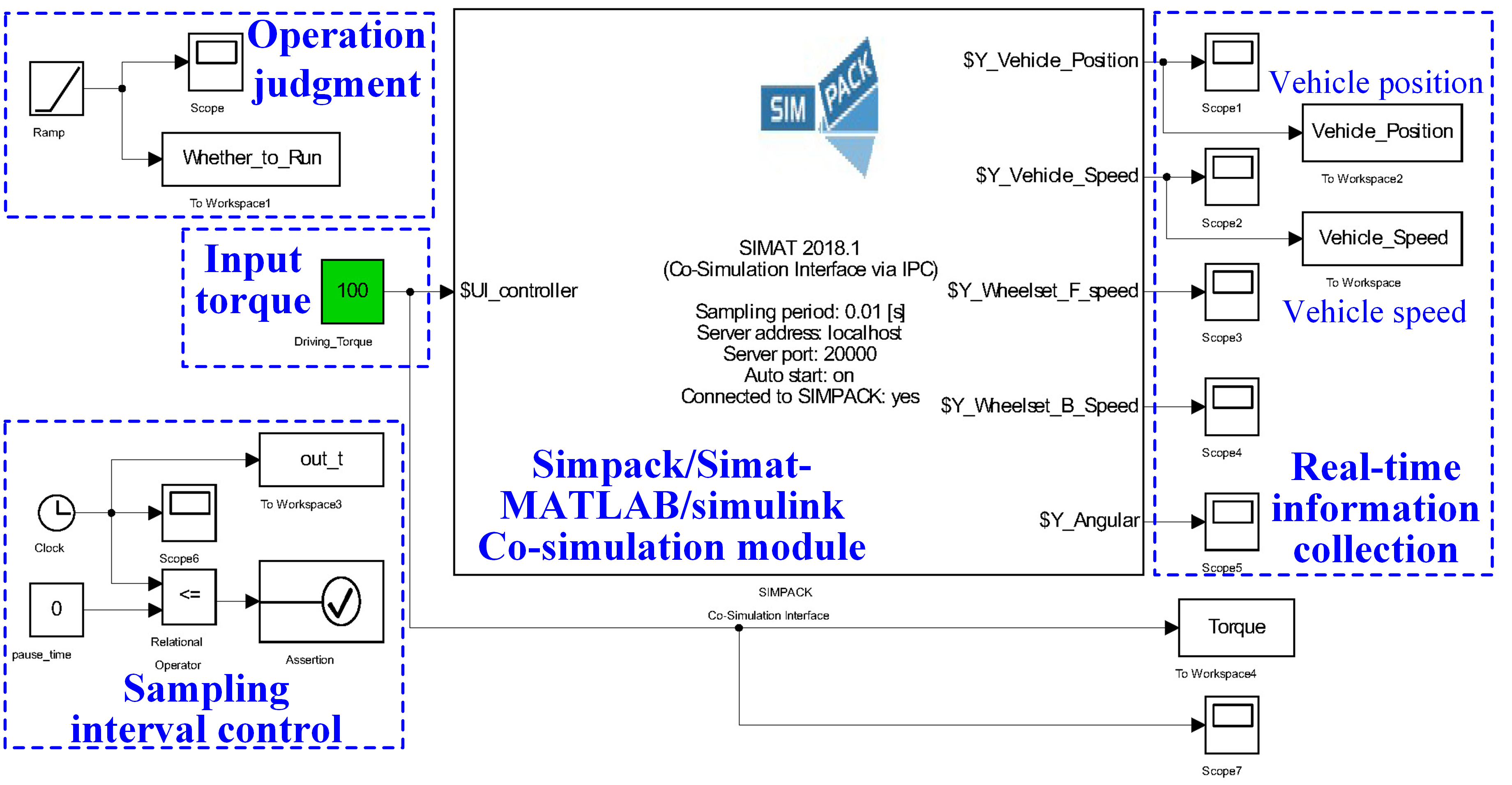}
\caption{Composition of co-simulation and verification platform for autonomous control of the mining electric locomotive.}
\label{fig:8}
\end{figure}

The software and hardware we used to build the co-simulation platform is shown in Table \ref{tab:table4}. As shown in Fig. \ref{fig:8}, the co-simulation platform for autonomous control of mining electric locomotives includes three parts: dynamic module, data analysis module, and control algorithm module.

\begin{table}[ht]
\small\sf\centering
\caption{Co-simulation Platform and Environment\label{tab:table4}.}
\begin{tabular}{ll}
\toprule
Category & Parameter\\
\midrule
\texttt{System} & Windows 11 Professional 64-bit\\
\texttt{CPU} & \makecell[l]{Intel(R) Core(TM) i9-9900K \\ CPU @ 3.60GHz}\\
\texttt{GPU} & NVIDIA GeForce RTX 2070\\
\texttt{RAM} & 32.0 GB\\
\texttt{Software} & \makecell[l]{PyCharm 2021; \\ MATLAB R2014b; \\ Simpack 2018}\\
\texttt{Program language} & Python 2.7\\
\texttt{3rd-party library} & \makecell[l]{matlabengineforpython R2014b; \\ openpyxl 2.6.4}\\
\bottomrule
\end{tabular}
\end{table}

During the simulation iteration process, especially in the application of intelligent algorithms, there will be error warnings issued by the dynamics software which will terminate the learning process. This is because the agent randomly generates actions at the beginning of the iteration. The value of random-generated actions may be too large or too small, which does not meet the normal operation logic of the dynamics software, causing the dynamics software to report an error. Therefore, in order to ensure the operational continuity of the multi software co-simulation platform, we set up a running judgment module in MATLAB/Simulink, adding a ramp signal in this module and reading the signal by a Python control algorithm. If the signal value is updated, MATLAB will run normally. Otherwise, the algorithm will skip to the next iteration and restart MATLAB.

In order to achieve dynamic simulation of autonomous mining electric locomotives, we also use the Simulink Assertion module to establish a sampling interval control module in MATLAB/Simulink. We set trigger logic in Assertion to enable Simulink to trigger simulation pauses after the variable sampling interval required by the control algorithm. We can control Simulink to continue running through the set\_param() command. The advantage of this step is that the sampling time of the control algorithm can be updated by adjusting the algorithm parameters, rather than must be consistent with the sampling time of Simpack and MATLAB. The control method is more flexible. The sampling time of Simpack and Simulink should be consistent to 0.001s. The sampling interval length required by the control algorithm can be greater than or equal to the sampling time length in Simulink based on the specific needs. In this study, the fixed step length is 0.01s.
\subsection{Simulation Setup and Algorithm Structure}
According to the requirements of JB/T 4091-2014 "Technical Conditions for Explosion-proof Special Battery Electric Locomotives in Coal Mines", the electric locomotive shall pass the specified minimum curve radius at a speed of 50\% on the straight track. The maximum speed limit of the mining electric locomotive driving on curves is 1.42 m/s, and the maximum speed limit driving on the straight track is 2.83 m/s. 

In the simulation, we simulate the speed of the mining electric locomotive on the straight and curved track, and the status of the vehicle when meeting the dynamic obstacles. We will also verify whether the mining electric locomotive can start and brake safely at the destination. In order to reduce the total simulation time, we set the obstacle occurrence position to be 5m in front of the starting point of the mining electric locomotive operation, and the speed of the obstacle is 0.01 m/s. The minimum distance between vehicles and obstacles is set to 5m default. Parameter settings in the simulation are only used to evaluate the algorithm and do not provide reference for the actual operation of the electric locomotive. However, parameters can be adjusted during underground operation to meet actual operation requirements.

In the early exploration stage of the iteration, input uncertainty can lead to Simpack crashing, which will further result in excessive reward values in the algorithm. The Algorithm \ref{alg:alg2} shows that how we design to eliminate extreme values in the simulation. The control process mainly includes$\colon$ 1) judging whether the Simulink operates normally, 2) obtaining the action values of the mining electric locomotive with IEG strategy and writing them into Simulink, 3) reading out the real-time status of the mining electric locomotive from Simulink and importing them into Excel, 4) eliminating extreme values and updating the Q-table. The parameter setting for the mining electric locomotive autonomous control algorithm are shown in Table \ref{tab:table5}. 

\begin{table}[ht]
\small\sf\centering
\caption{Parameter Setting for the Mining Electric Locomotive Autonomous Control Algorithm\label{tab:table5}.}
\begin{tabular}{lll}
\toprule
Hyperparameter & Value & Description\\
\midrule
\texttt{initial\_epsilon} & 0.01 & Initial exploration rate\\
\texttt{final\_epsilon} & 1 & Final exploration rate\\
\texttt{alpha} & 0.2 & Learning rate of Q-learning\\
\texttt{gamma} & 0.8 & Discount factor gamma\\
\texttt{sample\_time} & 0.01s & Sample time of the algorithm\\
\texttt{run\_time} & 45s & Running time of the algorithm\\
\texttt{max\_episodes} & 500 & Number of episodes\\
\bottomrule
\end{tabular}
\end{table}

\begin{algorithm}[!t]
\caption{RL for the mining electric locomotive controlling}\label{alg:alg2}
\begin{algorithmic}
\STATE 
\STATE {\textbf{Input:} V\_curr\_spd, V\_curr\_pos }
\STATE {\textbf{Output:} V\_driving\_torque } 
\STATE {\textbf{Initialize:} array}$_\text{reward\_sum},\varepsilon$
\STATE {\textbf{FOR}  episode in max\_episodes:} 
\STATE \hspace{0.5cm}{Start Simulink}
\STATE \hspace{0.5cm}{Generate the location of roadway obstacles randomly}
\STATE \hspace{0.5cm}{\textbf{Update}} $\varepsilon$
\STATE \hspace{0.5cm}{\textbf{Initialize:} reward\_sum, Is\_Simulink\_ok}
\STATE \hspace{0.5cm}{\textbf{Sample:}}
\STATE \hspace{1cm}{\textbf{FOR} step in max\_sampling\_number:}
\STATE \hspace{1.5cm}{\textbf{IF} Is\_Simulink\_ok == True:}
\STATE \hspace{2cm}{Get V\_action and V\_drving\_torque} 
\STATE \hspace{2cm}{\quad based on $\varepsilon${-greedy}}
\STATE \hspace{2cm}{Write V\_driving\_torque into Simulink}
\STATE \hspace{2cm}{Change Simulink running parameter}
\STATE \hspace{2cm}{Get V\_runtime, V\_curr\_spd, V\_curr\_pos}
\STATE \hspace{2cm}{\quad from Simulink}
\STATE \hspace{2cm}{Write V\_runtime, V\_curr\_spd, V\_curr\_pos}
\STATE \hspace{2cm}{\quad into Excel}
\STATE \hspace{2cm}{Get obstacle\_pos and Write obstacle\_pos}
\STATE \hspace{2cm}{\quad into Excel}
\STATE \hspace{2cm}{Get V\_new\_state}
\STATE \hspace{2cm}{\textbf{IF} step} $\leq 5:$ 
\STATE \hspace{2.5cm}{V\_reward=0}
\STATE \hspace{2cm}{\textbf{ELSE:}} 
\STATE \hspace{2.5cm}{Get V\_reward}
\STATE \hspace{2cm}{\textbf{IF} Is\_V\_reach\_dest == False:}
\STATE \hspace{2.5cm}{\textbf{Update Q-table}}
\STATE \hspace{2cm}{\textbf{ELSE:}}
\STATE \hspace{2.5cm}{\textbf{Update Q-table}}
\STATE \hspace{2.5cm}{break}
\STATE \hspace{2cm}{\textbf{IF} Is\_V\_draw\_back and V\_curr\_pos} $<$ {-0.05:}
\STATE \hspace{2.5cm}{break}
\STATE \hspace{2cm}{\textbf{IF} Is\_Simpack\_error == True:}
\STATE \hspace{2.5cm}{break}
\STATE \hspace{2cm}{\textbf{IF} Is\_V\_crash\_obstacle and V\_curr\_pos==0:}
\STATE \hspace{2.5cm}{break}
\STATE \hspace{2cm}{reward\_sum += V\_reward }
\STATE \hspace{2cm}{reward\_avg = reward\_sum/step}
\STATE \hspace{2cm}{V\_state = V\_new\_state}
\STATE \hspace{1.5cm}{\textbf{ELSE:}}
\STATE \hspace{2cm}{break}
\STATE \hspace{0.5cm}{\textbf{Update} array}$_\text{reward\_sum}$ and Write array$_\text{reward\_sum}$
\STATE \hspace{0.5cm}{\quad into Excel}
\STATE \hspace{0.5cm}{Terminate Simulink}
\STATE \hspace{0.5cm}{Save Excel}
\end{algorithmic}
\label{alg2}
\end{algorithm}

\subsection{Simulation Results}
\begin{figure}[!t]
\centering
\includegraphics[width=\linewidth]{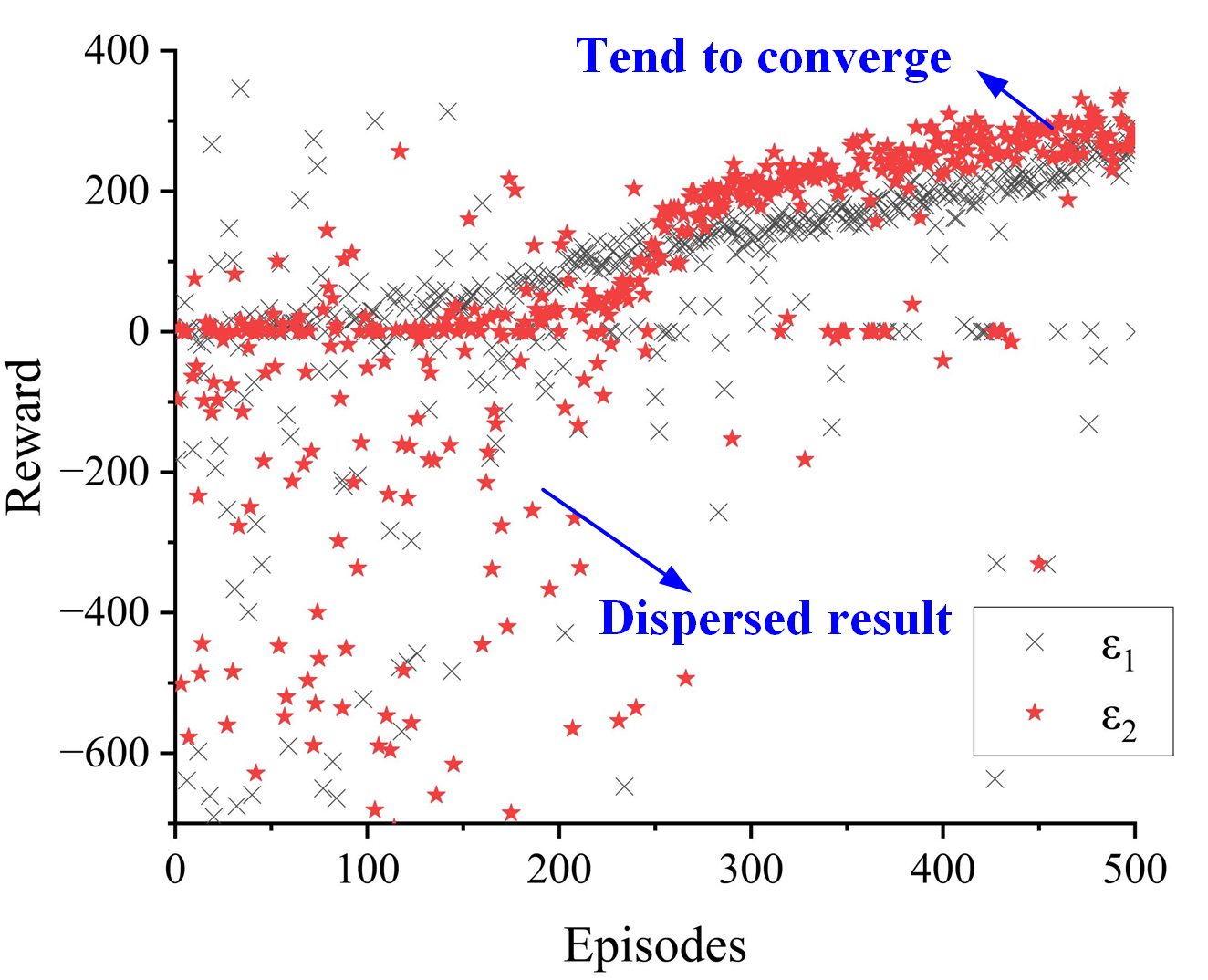}
\caption{Results of reward with different values of $\varepsilon$.}
\label{fig:9}
\end{figure}

After all iterative steps, we can obtain the results of reward value shown in Fig. \ref{fig:9}. With the number of iterations increasing, the learning process with an exploration rate of $\varepsilon_{1}$ can achieve an approximate linear increase in the reward value. However, the final reward value obtained by this learning process is smaller than that with IEG (an exploration rate of $\varepsilon_{2}$). In the early stage of the algorithm iteration, the two algorithms gain some discrete and poor results due to the more emphasis on trying different actions during the exploration process. In the later stage, the results of the algorithm obtain a relatively ideal range of reward values. This is because early attempts were made to bring different benefits from more actions. This will provide agents with more reliable experience values that can be utilized in the later stage. We can clearly find that when the learning process proceeds to 460-500 iterations with IEG, the reward value results gradually converge within a constant reward value range, and the control method is more reliable. At this stage, the rewards obtained by the learning process with IEG are approximately 24.3\% higher than those obtained by the learning process with an exploration rate of $\varepsilon_{1}$.

\begin{figure}[!t]
\centering
\includegraphics[width=\linewidth]{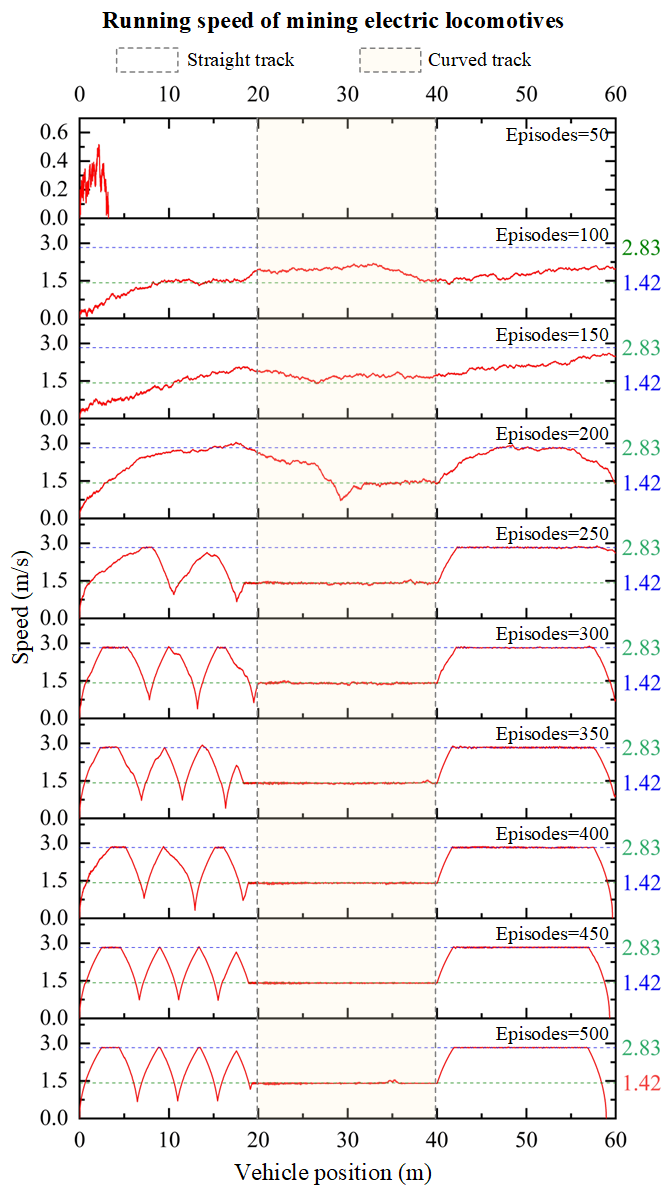}
\caption{The speed of the mining electric locomotive in different track sections (per 50 iterations).}
\label{fig:10}
\end{figure}

Due to the large amount of data, we take the results of every 50 iterations to draw the following velocity position curve, which is shown in Fig. \ref{fig:10}. At the beginning of the algorithm iteration, the autonomous mining electric locomotive cannot reach the maximum speed of the corresponding track section, and the driving speed fluctuates greatly. At this stage, the exploration rate of the mining electric locomotive learning process is large, and it is prone to errors in the operation of the dynamic model caused by too large/too small input action values, resulting in error results. However, as the algorithm iterates, the agent gradually improves its operational efficiency within the speed limit of the track section, and the speed changes are relatively stable. At the same time, in the middle and later stages of the learning process, the speed of the mining electric locomotive will be adjusted due to the emergence of dynamic obstacles, and it will decrease to zero at the destination position.

In order to evaluate the effect of maintaining the set safety distance between the electric locomotive and the dynamic obstacle, we take the closest distance between the vehicle and the obstacle in front at the moment during each iteration process, and define the difference between this distance and the safety distance $s_\text{safe}$ as safety evaluation index for car following, as shown in equation:
\begin{equation}
\label{deqn_9}
\begin{split}
s(t)={\min\limits_{episode \in [1, max\_episodes]}} {((p_{\text{obstacle},t}-p_{\text{vehicle},t})-s_\text{safe})}
\end{split}
\end{equation}
where $p_{\text{obstacle},t}$ and $p_{\text{vehicle},t}$  are the position of the obstacle and of the vehicle at the moment $t$.
Obviously, when the evaluation index is equal to 0, the vehicle can always maintain a set safe distance from dynamic obstacles, which is the ideal result we want. When the evaluation index is less than 0, it indicates that the minimum distance that the vehicle can maintain from dynamic obstacles is less than the set safety distance, and the closer the evaluation index is to 0, the better the control effect. 

\begin{figure}[!t]
\centering
\includegraphics[width=\linewidth]{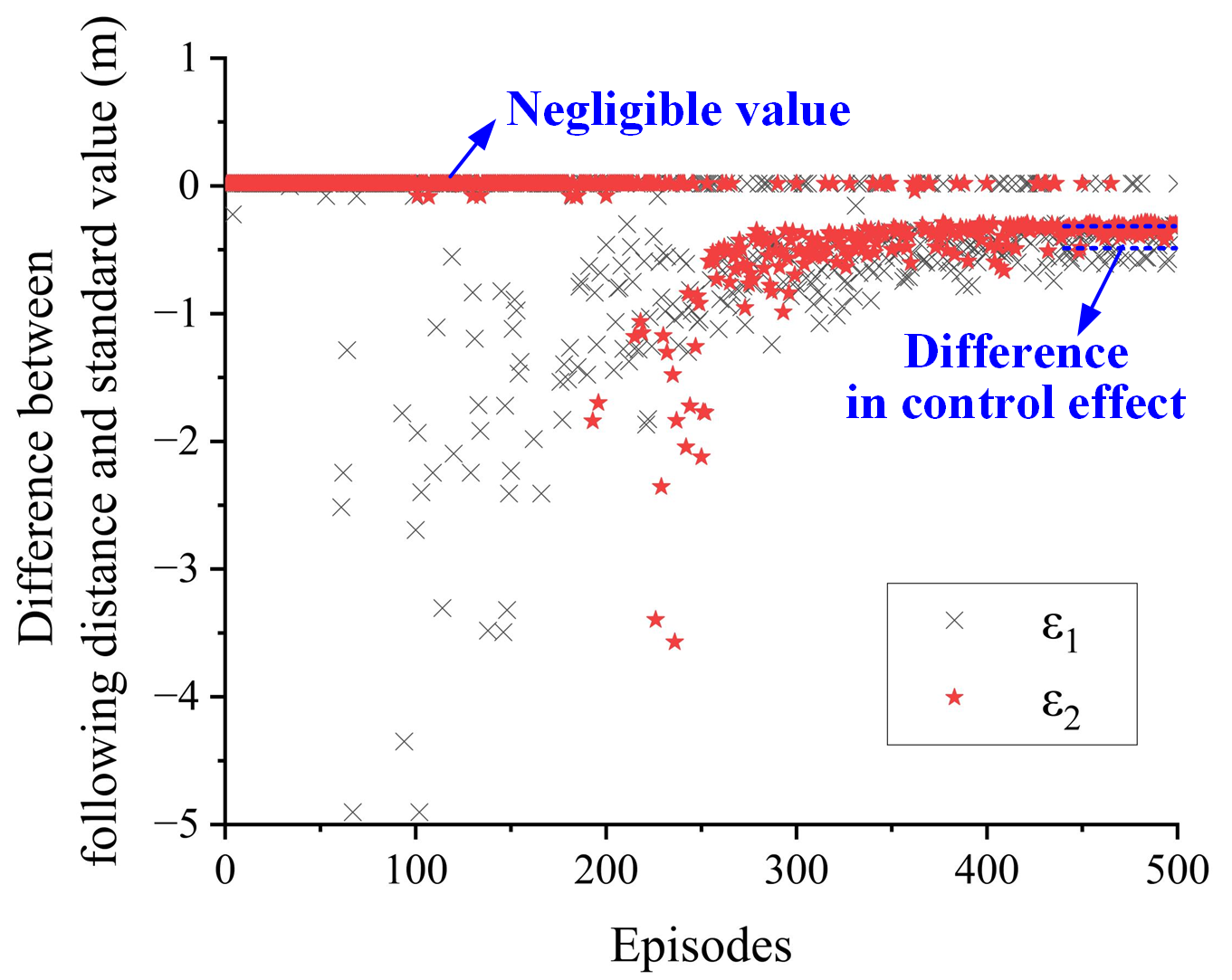}
\caption{Safety evaluation index for car following with different values of $\varepsilon$.}
\label{fig:11}
\end{figure}

The safety evaluation results are shown in Fig. \ref{fig:11}. In the early stages of the algorithm iteration, the agent pay more attention to exploring unknown processes, resulting in poor learning effectiveness. This phenomenon is more obvious when using IEG. Due to the poor learning effect, the vehicle speed always fails to reach the set maximum value, so it can always maintain a large distance from the obstacles. Therefore, the evaluation indicators can be ignored for evaluating the control effect of the algorithm at the early stage of the algorithm iteration.

In the middle and later stages of the algorithm iteration, the vehicle gradually obtains a preliminary autonomous control strategy. The effect of maintaining a safe distance from obstacles in a learning process with IEG is better than that of a learning process with an exploration rate of $\varepsilon_{1}$. 

In terms of the control effect of maintaining a safe distance between the electric locomotive and the obstacle, the learning process with IEG is more stable in the later stage of iteration. The learning process with an exploration rate of $\varepsilon_{1}$ fluctuates greatly. The learning effect of the agent in maintaining a safe distance from the obstacle is more reliable when we adopt IEG. 

Taking a 441-500 iteration process where the car following evaluation index tends to converge, we can calculate that the evaluation index of the learning process with IEG is about 19.01\% higher than that of the learning process with an exploration rate of $\varepsilon_{1}$. It means that the vehicle can operate more safely when there are dynamic obstacles in front of it.

\section{Conclusion}
This paper applied Reinforcement Learning into automatic control of mining electric locomotives. 
We proposed an IEG strategy, a reinforcement learning algorithm which allow controlling the mining electric locomotives automatically in different complex mining environments.
In the design of the reward function of IEG, we considered all influencing factors (driving speed of the mining electric locomotives, speed limit on different driving track sections, distance between electric locomotives, dynamic obstacles) so that mining electric locomotives can follow the vehicle in front safely and avoid the obstacles timely. 
We also built the first industrial multi-software co-simulation platform based on Simpack, MATLAB/Simulink and Python, which can realize closed loop testing of automatic control algorithm of mining electric locomotives.
The  simulation results show that a 24.3\% increase in the reward value and a 19.01\% increase in the following evaluation index, compared to the traditional linear decreasing $\varepsilon$-greedy strategy. The results demonstrate that the relationship between exploration and exploitation of $\varepsilon$-greedy strategy is well balanced, and the autonomous control method of mining electric locomotives is feasible and effective.

\begin{acks}
This work was supported by National Key R\&D Program 2020YFB1314100, China.
\end{acks}

\bibliographystyle{SageV}

\end{document}